\documentclass[amsmath,aps,showpacs,a4paper,10pt]{revtex4}
 \usepackage{epsf}
 \usepackage{graphicx}    % Include figure files

\begin{document}

 \newcommand{\be}[1]{\begin{equation}\label{#1}}
 \newcommand{\ee}{\end{equation}}
 \newcommand{\bea}{\begin{eqnarray}}
 \newcommand{\eea}{\end{eqnarray}}
 \def\disp{\displaystyle}

 \def\gsim{ \lower .75ex \hbox{$\sim$} \llap{\raise .27ex \hbox{$>$}} }
 \def\lsim{ \lower .75ex \hbox{$\sim$} \llap{\raise .27ex \hbox{$<$}} }

 \begin{titlepage}

 \begin{flushright}
 arXiv:1204.4032
 \end{flushright}

 \title{\Large \bf Pilgrim Dark Energy}

 \author{Hao~Wei\,}
 \email[\,email address:\ ]{haowei@bit.edu.cn}
 \affiliation{School of Physics, Beijing Institute
 of Technology, Beijing 100081, China}

 \begin{abstract}\vspace{1cm}
 \centerline{\bf ABSTRACT}\vspace{2mm}
 In the present work, we reconsider the idea of holographic
 dark energy. One of its key points is the formation of the
 black hole. And then, we propose the so-called ``pilgrim dark
 energy'' based on the speculation that the repulsive force
 contributed by the phantom-like dark energy ($w<-1$) is
 strong enough to prevent the formation of the black hole. We
 also consider the cosmological constraints on pilgrim dark
 energy by using the latest observational data. Of course,
 one can instead regard pilgrim dark energy as a purely
 phenomenological model without any physical motivation. We
 also briefly discuss this issue.
 \end{abstract}


 \maketitle

 \end{titlepage}

 \renewcommand{\baselinestretch}{1.1}

%============================= section 1 ===================================

\section{Introduction}\label{sec1}

Since the discovery of the current accelerated expansion of
 our universe, dark energy has been one of the most active
 fields in physics and astronomy~\cite{r1}. Of course,
 the simplest candidate of dark energy is a tiny positive
 cosmological constant, $\Lambda$. However, as is well known,
 it is plagued with the cosmological constant problem (the
 fine-tuning problem) and the cosmological coincidence
 problem. Therefore, many alternative dark energy models have
 been proposed.

Recently, the so-called holographic dark energy (HDE) has been
 considered as an interesting candidate of dark energy, which
 has been studied extensively in the literature. For a quantum
 gravity system, the local quantum field cannot contain too
 many degrees of freedom, otherwise the black hole forms and
 then the quantum field theory breaks down. In the black hole
 thermodynamics~\cite{r2,r3}, there is a maximum entropy in a
 box of size $L$, namely the Bekenstein-Hawking entropy bound
 $S_{BH}$, which scales as the area of the box $\sim L^2$,
 rather than the volume $\sim L^3$. To avoid the breakdown of
 the local quantum field theory, Cohen~{\it et al.}~\cite{r4}
 proposed the so-called energy bound, which is more restrictive
 than the entropy bound. If $\rho_\Lambda$ is the quantum
 zero-point energy density caused by a short distance cut-off,
 the total energy in a box of size $L$ cannot exceed the mass
 of a black hole of the same size~\cite{r4},
 namely $\rho_\Lambda L^3\,\lsim\, m_p^2 L$, where
 $m_p\equiv (8\pi G)^{-1/2}$ is the reduced Planck mass. The
 largest IR cut-off $L$ is the one saturating the inequality.
 Therefore, one has
 \be{eq1}
 \rho_\Lambda=3c^2 m_p^2 L^{-2}\,,
 \ee
 where the numerical constant $3c^2$ is introduced for
 convenience. In the literature, many IR cut-offs $L$ have been
 considered, such as the Hubble horizon $H^{-1}$~\cite{r5,r6},
 the particle horizon $R_H\equiv a\int_0^t d\tilde{t}/a
 =a\int_0^a d\tilde{a}/(H\tilde{a}^2)$~\cite{r7,r43}, the
 future event horizon $R_h\equiv a\int_t^\infty d\tilde{t}/a
 =a\int_a^\infty d\tilde{a}/(H\tilde{a}^2)$~\cite{r8}, the
 Ricci scalar curvature radius, which is actually proportional
 to the causal connection scale of perturbations in the flat
 universe $R_{\rm CC}=(\dot{H}+2H^2)^{-1/2}$~\cite{r9}, the
 formal generalization of $R_{\rm CC}$, namely
 $(\alpha H^2+\beta\dot{H})^{-1/2}$~\cite{r10}, the age of our
 universe $T=\int_0^a d\tilde{a}/(H\tilde{a})$~\cite{r11},
 the conformal age of our universe $\eta\equiv\int_0^t d\tilde{t}/a
 =\int_0^a d\tilde{a}/(\tilde{a}^2 H)$~\cite{r12}, the radius
 of the cosmic null hypersurface~\cite{r13}, the so-called
 conformal-age-like length~\cite{r14}, where $H\equiv\dot{a}/a$
 is the Hubble parameter; $a=(1+z)^{-1}$ is the scale factor
 of our universe; $z$ is the redshift; we have set $a_0=1$;
 the subscript ``0'' indicates the present value of the
 corresponding quantity; a dot denotes the derivative with
 respect to cosmic time $t$.

In the present work, we reconsider the idea of holographic dark
 energy. One of its key points is the formation of the black
 hole. If we can prevent the formation of the black hole, the
 energy bound proposed by Cohen~{\it et al.}~\cite{r4} could be
 violated. If the repulsive force is strong enough, it might
 resist the matter collapse and then black hole does not
 form. So, what can contribute the strong repulsive force?
 Nowadays, it is well known that dark energy can contribute a
 repulsive force, since its equation-of-state parameter~(EoS)
 $w<-1/3$. However, we will see below that the repulsive force
 contributed by the quintessence-like dark energy ($w>-1$) is
 not strong enough to prevent the formation of the black hole.
 Therefore, we focus on the phantom-like dark energy ($w<-1$).
 Firstly, it is well known that everything will be completely
 torn~up before our universe ends in the big rip caused by the
 phantom-like dark energy. Even the black hole will also be
 completely torn~up. So, one can find that the repulsive force
 contributed by the phantom-like dark energy is strong enough
 to destroy the black hole. Secondly, in the famous work
 by Babichev~{\it et al.}~\cite{r15}, it is shown that
 accretion of phantom-like dark energy is accompanied with the
 gradual decrease of the black hole mass. Of course, this issue
 has not been completely settled by now. In~\cite{r16},
 Gao~{\it et al.} claimed that the physical black hole mass may
 instead increase due to the accretion of phantom energy, but
 the Cosmic Censorship Conjecture will be violated. Later,
 Gonzalez and Guzman~\cite{r17} presented the full non-linear
 study of phantom scalar field accreted into a black hole, and
 found that the accretion of the phantom scalar field into the
 black hole can reduce its area down to $50\%$ within time
 scales of the order of few masses of the initial horizon.
 In~\cite{r18}, Sun also claimed that in the phantom dark
 energy universe the black hole mass becomes zero before the
 big rip is reached. On the other hand, Jamil and
 Qadir~\cite{r19} claimed that the phantom energy accretion
 contribute to a decrease of the mass of the primordial black
 hole. Recently, Sharif and Abbas~\cite{r20} also found that
 mass of the black hole decreases due to phantom accretion.
 Thirdly, Harada~{\it et al.}~\cite{r21} claimed that there
 is no self-similar black hole solution in an universe with a
 stiff fluid or scalar field or quintessence. It is worth
 noting that Akhoury~{\it et al.}~\cite{r44} independently
 obtained the similar result from a very different perspective.
 Chapline~\cite{r22} claimed that black holes might not~exist
 in our real world, because the negative pressure contributed
 by dark energy might prevent black holes from forming. Note
 that in~\cite{r21,r22,r44} they only require dark energy
 violating the strong energy condition. Of course, they are
 wrong, because the repulsive force contributed by the
 quintessence-like dark energy ($w>-1$) is not strong enough to
 prevent the formation of the black hole. This point can be
 seen in~\cite{r23}, where by considering some specific
 McVittie solutions, Li and Wang showed that black holes can
 exist in a Friedmann-Robertson-Walker~(FRW) universe dominated
 by dark energy. However, they require that the weak energy
 condition should be satisfied (although the other three energy
 conditions can be violated). Here, we would like to stress
 that the phantom-like dark energy does not satisfy the weak
 energy condition (in fact it violates all energy conditions).
 Therefore, the conclusion of Li and Wang~\cite{r23} cannot be
 applied to the phantom-like dark energy ($w<-1$). On the other
 hand, Rahaman~{\it et~al.}~\cite{r24} claimed that they have
 found an exact solution of spherically symmetrical Einstein
 equations describing a black hole with a special type
 ``phantom'' energy source. Unfortunately, this special type
 ``phantom'' energy source used in~\cite{r24} is characterized
 by $p=-\rho$ in fact. Obviously, it is not the commonly
 called phantom energy ($p<-\rho$), and hence the
 conclusion of~\cite{r24} cannot be applied here.

Together with these three arguments mentioned above, it is
 reasonable to speculate that the repulsive force contributed
 by the phantom-like dark energy ($w<-1$) is strong enough to
 prevent the formation of the black hole. Of course, we admit
 that this speculation need further and solid proofs which are
 still absent. Anyway, we suggest considering what will happen
 to the argument of holographic dark energy if this speculation
 is true. In Sec.~\ref{sec2}, we propose the
 so-called ``pilgrim dark energy'' based on this speculation.
 Then, in Sec.~\ref{sec3}, we consider the cosmological
 constraints on pilgrim dark energy by using the latest
 observational data. Some concluding remarks are given in
 Sec.~\ref{sec4}.

%============================= section 2 ===================================

\section{Pilgrim dark energy}\label{sec2}

Here, we propose the so-called pilgrim dark energy (PDE) based
 on the aforementioned speculation that the repulsive force
 contributed by the phantom-like dark energy ($w<-1$) is strong
 enough to prevent the formation of the black hole. If this
 speculation is true, the energy bound proposed
 by Cohen~{\it et~al.}~\cite{r4} could be violated, namely
 the total energy in a box of size $L$ could exceed the
 mass of a black hole of the same size~\cite{r4}, i.e.,
 $\rho_\Lambda L^3\,\gsim\, m_p^2 L\,$. Therefore, the first
 property of pilgrim dark energy is
 \be{eq2}
 \rho_\Lambda\,\gsim\, m_p^2 L^{-2}\,.
 \ee
 To implement Eq.~(\ref{eq2}), the simplest way is to set
 \be{eq3}
 \rho_\Lambda=3n^2 m_p^{4-s}L^{-s}\,,
 \ee
 where $n$ and $s$ are both dimensionless constants. Thus,
 from Eqs.~(\ref{eq2}) and (\ref{eq3}),
 $\rho_\Lambda\sim m_p^{4-s}L^{-s}\,\gsim\, m_p^2 L^{-2}$
 leads to $L^{2-s}\,\gsim\, m_p^{s-2}=\ell_p^{2-s}$, where
 $\ell_p=m_p^{-1}=1.616\times 10^{-33}\,{\rm cm}$ is the
 reduced Planck length, which is extremely short length
 in fact. Obviously, since $L>\ell_p$ in general,
 it is required that
 \be{eq4}
 s\leq 2\,.
 \ee
 As mentioned above, the second requirement of pilgrim
 dark energy is to be phantom-like, namely
 \be{eq5}
 w_\Lambda<-1\,.
 \ee
 Here, in order to obtain the EoS of pilgrim dark energy, we
 should choose a specific cut-off $L$. Of course, the simplest
 choice is the Hubble horizon $L=H^{-1}$. In the epoch
 dominated by pilgrim dark energy, the Friedmann equation
 $H^2\to\rho_\Lambda/(3m_p^2)=n^2 m_p^{2-s}H^s$ leads to
 $H\to const.$, which corresponds to $w_\Lambda\to -1$.
 Therefore, our universe will end in a de Sitter phase, rather
 than big rip. In the matter-dominated epoch,
 $H^2\sim\rho_m\sim a^{-3}$. So,
 $\rho_\Lambda\sim H^s\sim a^{-3s/2}$. Thus, we have
 $w_\Lambda=-1+s/2$. From Eq.~(\ref{eq5}), $s<0$ is required.
 Let us consider the general case. Substituting
 $\rho_\Lambda\propto H^s$ into the energy conservation equation
 $\dot{\rho}_\Lambda+3H\rho_\Lambda\left(1+w_\Lambda\right)=0$,
 we have
 \be{eq6}
 w_\Lambda=-1-\frac{s\dot{H}}{3H^2}\,.
 \ee
 One might naively consider that $\dot{H}>0$ in the late time,
 since $w_\Lambda<-1$. However, it is wrong. The total
 EoS $w_{tot}=\Omega_\Lambda w_\Lambda\to -1$,
 where $\Omega_\Lambda$ is the fractional energy density
 of pilgrim dark energy. Therefore, $\dot{H}<0$ instead holds
 in the late time. If $\dot{H}<0$ is valid in the whole
 cosmic history, from Eqs.~(\ref{eq5}) and~(\ref{eq6}), it
 is required that
 \be{eq7}
 s<0\,.
 \ee
 So, there is no conflict between the matter-dominated epoch
 and the late time. $w_\Lambda$ starts from $-1+s/2$ in the
 matter-dominated epoch and goes asymptotically to $-1$ in
 the late time; $w_\Lambda<-1$ always and it never crosses
 the phantom divide $w=-1$ in the whole cosmic history. To
 be clearer, let us see a particular example of $s=-2$.
 In this case, substituting $\rho_m=\rho_{m0}a^{-3}$ and
 Eq.~(\ref{eq3}) with $L=H^{-1}$ into the Friedmann equation,
 we obtain
 \be{eq8}
 H^2=\frac{1}{3m_p^2}\left(\rho_\Lambda+\rho_m\right)=
 n^2 m_p^4 H^{-2}+\frac{\rho_{m0}}{3m_p^2}\,a^{-3}
 \equiv c_1\,H^{-2}+c_2\,a^{-3}\,.
 \ee
 Note that we consider a flat FRW universe which only contains
 pressureless matter and dark energy throughout this work.
 Requiring $H^2\geq 0$, we can solve Eq.~(\ref{eq8}) and obtain
 \be{eq9}
 H^2=\frac{1}{2}\left(c_2\,a^{-3}+
 \sqrt{4c_1+c_2^2\,a^{-6}}\,\right).
 \ee
 When $a\to 0$, $H^2\to c_2\,a^{-3}$, which coincides with
 the one of matter-dominated epoch. When $a\to\infty$,
 $H^2\to\sqrt{c_1}=const.$, which coincides with the one of
 de Sitter phase in the late time. From Eq.~(\ref{eq9}),
 it is easy to see that $H$ decreases when $a$ increases;
 $H$ is a monotonically decreasing function of $a$. Thus,
 $\dot{H}<0$ indeed holds in the whole cosmic history. So, if
 $s<0$, we have $w_\Lambda<-1$ in the whole cosmic history.
 Note that if $s=0$, from Eq.~(\ref{eq3}), it is easy to see
 that our pilgrim dark energy model reduces to the well-known
 $\Lambda$CDM model. Considering this point as well as
 Eqs.~(\ref{eq4}) and~(\ref{eq7}), we allow
 \be{eq10}
 s\leq 0\,,
 \ee
 in order to satisfy the requirements (\ref{eq2})
 and (\ref{eq5}), and include $\Lambda$CDM as a special
 case (although $w_\Lambda=-1$ in this case).

At first glance, one might consider that pilgrim dark energy
 model is a three-parameter model, which contains $n$, $s$ and
 $\Omega_{m0}$ as model parameters (where $\Omega_{m0}$ is the
 present fractional energy density of pressureless matter).
 However, one of them is not independent in fact. Substituting
 $\rho_m=\rho_{m0}a^{-3}$ and Eq.~(\ref{eq3}) with $L=H^{-1}$
 into the Friedmann equation, namely
 $3m_p^2 H^2=\rho_\Lambda+\rho_m$, we have
 \be{eq11}
 n^2 m_p^{2-s}H^{s-2}+\frac{\rho_{m0}}{3m_p^2 H^2}\,a^{-3}=1\,.
 \ee
 Introducing dimensionless $E\equiv H/H_0$
 and $\tilde{n}^2\equiv n^2 (m_p/H_0)^{2-s}$, we recast
 Eq.~(\ref{eq11}) as
 \be{eq12}
 \tilde{n}^2 E^{s-2}+\Omega_{m0}E^{-2}a^{-3}=1\,.
 \ee
 Requiring $E(a=1)=1$ by definition, we find that
 \be{eq13}
 \tilde{n}^2=1-\Omega_{m0}\,.
 \ee
 So, $n$ (or equivalently $\tilde{n}$) is not independent.
 Finally, the Friedmann equation becomes
 \be{eq14}
 \left(1-\Omega_{m0}\right) E^{s-2}+\Omega_{m0} E^{-2} (1+z)^3=1\,.
 \ee
 Obviously, there are only two free model parameters, namely
 $\Omega_{m0}$ and $s$. From Eq.~(\ref{eq14}), one can obtain
 $E(z)$ as a function of redshift $z$, if model parameters
 $\Omega_{m0}$ and $s$ are given. Note that $E$ is a real
 number and $E\geq 0$ is required by definition.

Here, we would like to say some words before going further.
 As is shown above, pilgrim dark energy is based
 on the aforementioned speculation that the repulsive force
 contributed by the phantom-like dark energy ($w<-1$) is
 strong enough to prevent the formation of the black hole.
 In fact, this speculation is proposed just from
 some arguments, and it has no solid foundation so far. If
 one cannot agree the physical motivation of pilgrim dark
 energy presented here, we suggest that one can instead
 regard pilgrim dark energy as a purely phenomenological
 model without invoking any physical motivation.

%============================= Fig. 1 =================================

 \begin{center}
 \begin{figure}[tbhp]
 \centering
 \includegraphics[width=0.5\textwidth]{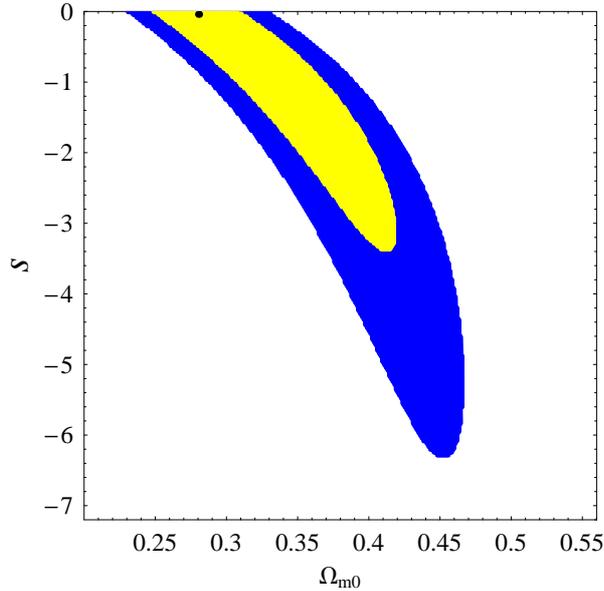}
 \caption{\label{fig1}
 The $68.3\%$ and $95.4\%$ confidence level contours in the
 $\Omega_{m0}-s$ parameter space. The best-fit parameters
 are also indicated by a black solid point. This result is
 obtained by using the data of 580 Union2.1 SNIa alone.}
 \end{figure}
 \end{center}

%======================================================================

\vspace{-10mm} % used here just for a more comfortable typesetting

%============================= section 3 ===================================

\section{Cosmological constraints on pilgrim dark energy}\label{sec3}

In this section, we consider the cosmological constraints on
 pilgrim dark energy by using the latest observational data.

At first, we use the observational data of Type~Ia supernovae
 (SNIa) alone. Recently, the Supernova Cosmology Project (SCP)
 collaboration released the updated Union2.1 compilation
 which consists of 580 SNIa~\cite{r25}. The Union2.1
 compilation is the largest published and spectroscopically
 confirmed SNIa sample to date. The data points of the 580
 Union2.1 SNIa compiled in~\cite{r25} are given in terms of
 the distance modulus $\mu_{obs}(z_i)$. On the other hand, the
 theoretical distance modulus is defined~as
 \be{eq15}
 \mu_{th}(z_i)\equiv 5\log_{10}D_L(z_i)+\mu_0\,,
 \ee
 where $\mu_0\equiv 42.38-5\log_{10}h$ and $h$ is the Hubble
 constant $H_0$ in units of $100~{\rm km/s/Mpc}$, whereas
 \be{eq16}
 D_L(z)=(1+z)\int_0^z \frac{d\tilde{z}}{E(\tilde{z};{\bf p})}\,,
 \ee
 in which ${\bf p}$ denotes the model parameters. The $\chi^2$
 from 580 Union2.1 SNIa is given by
 \be{eq17}
 \chi^2_{\mu}({\bf p})=\sum\limits_{i}\frac{\left[
 \mu_{obs}(z_i)-\mu_{th}(z_i)\right]^2}{\sigma^2(z_i)}\,,
 \ee
 where $\sigma$ is the corresponding $1\sigma$ error. The parameter
 $\mu_0$ is a nuisance parameter, but it is independent of the data
 points. One can perform an uniform marginalization over $\mu_0$.
 However, there is an alternative way. Following~\cite{r26,r27}, the
 minimization with respect to $\mu_0$ can be made by expanding the
 $\chi^2_{\mu}$ of Eq.~(\ref{eq17}) with respect to $\mu_0$ as
 \be{eq18}
 \chi^2_{\mu}({\bf p})=\tilde{A}-2\mu_0\tilde{B}+\mu_0^2\tilde{C}\,,
 \ee
 where
 $$\tilde{A}({\bf p})=\sum\limits_{i}\frac{\left[\mu_{obs}(z_i)
 -\mu_{th}(z_i;\mu_0=0,{\bf p})\right]^2}
 {\sigma_{\mu_{obs}}^2(z_i)}\,,$$
 $$\tilde{B}({\bf p})=\sum\limits_{i}\frac{\mu_{obs}(z_i)
 -\mu_{th}(z_i;\mu_0=0,{\bf p})}{\sigma_{\mu_{obs}}^2(z_i)}\,,
 ~~~~~~~~~~
 \tilde{C}=\sum\limits_{i}\frac{1}{\sigma_{\mu_{obs}}^2(z_i)}\,.$$
 Eq.~(\ref{eq18}) has a minimum for
 $\mu_0=\tilde{B}/\tilde{C}$ at
 \be{eq19}
 \tilde{\chi}^2_{\mu}({\bf p})=
 \tilde{A}({\bf p})-\frac{\tilde{B}({\bf p})^2}{\tilde{C}}\,.
 \ee
 Since $\chi^2_{\mu,\,min}=\tilde{\chi}^2_{\mu,\,min}$
 obviously (up to a constant), we can instead minimize
 $\tilde{\chi}^2_{\mu}$ which is independent of $\mu_0$. The
 best-fit model parameters are determined by minimizing the
 total $\chi^2$. When SNIa is used alone, we have
 $\chi^2=\tilde{\chi}^2_\mu$ which is given
 in Eq.~(\ref{eq19}). As in~\cite{r27,r28}, the $68.3\%$
 confidence level (C.L.) is determined by
 $\Delta\chi^2\equiv\chi^2-\chi^2_{min}\leq 1.0$, $2.3$ and
 $3.53$ for $n_p=1$, $2$ and $3$, respectively, where $n_p$ is
 the number of free model parameters. Similarly, the $95.4\%$
 confidence level is determined by
 $\Delta\chi^2\equiv\chi^2-\chi^2_{min}\leq 4.0$, $6.17$ and
 $8.02$ for $n_p=1$, $2$ and $3$, respectively. Here, we scan
 the $\Omega_{m0}-s$ parameter space (note that as mentioned
 above, $s\leq 0$ and $0\leq\Omega_{m0}\leq 1$ are required),
 and solve Eq.~(\ref{eq14}) to obtain $E(z)$ as a function of
 redshift $z$. Therefore, the corresponding $\chi^2$ is on
 hand. Finally, we find that the best fit has
 $\chi^2_{min}=562.226$, and the corresponding
 best-fit parameters are $\Omega_{m0}=0.280$ and $s=-0.04$.
 In Fig.~\ref{fig1}, we present the corresponding $68.3\%$ and
 $95.4\%$~C.L. contours in the $\Omega_{m0}-s$ parameter space.
 It is easy to see that although the best-fit $s$ is close to
 zero, there is a very big room for a significantly non-zero
 $s$ in the $95.4\%$ C.L. region. In fact, the viable $s$
 can extend to about $-6.4$ at $95.4\%$~C.L., or $-3.4$ at
 $68.3\%$~C.L. The price to have a significantly non-zero $s$
 is a larger $\Omega_{m0}$.

%============================= Fig. 2 =================================

 \begin{center}
 \begin{figure}[b]
 \centering
 \includegraphics[width=0.5\textwidth]{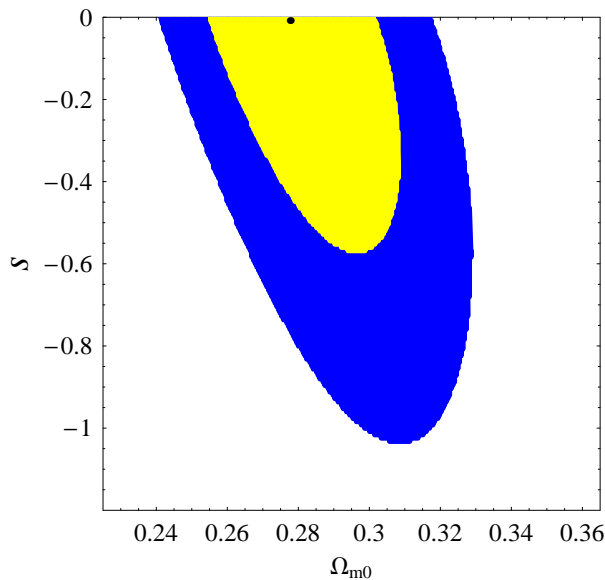}
 \caption{\label{fig2}
 The same as in Fig.~\ref{fig1}, except that this result is
 obtained by using the combined observational data of 580
 Union2.1 SNIa and the distance parameter $A$ from BAO.}
 \end{figure}
 \end{center}

%======================================================================

\vspace{-9.5mm} % used here just for a more comfortable typesetting

Next, we add the data from the observation of the large-scale
 structure (LSS). Here we use the distance parameter $A$ of
 the measurement of the baryon acoustic oscillation (BAO)
 peak in the distribution of SDSS luminous red
 galaxies~\cite{r29,r30}, which contains the main information
 of the observations of LSS. The distance parameter $A$ is
 given by
 \be{eq20}
 A\equiv\Omega_{m0}^{1/2}\,E(z_b)^{-1/3}\left[\frac{1}{z_b}
 \int_0^{z_b}\frac{d\tilde{z}}{E(\tilde{z})}\right]^{2/3}\,,
 \ee
 where $z_b=0.35$. In~\cite{r30}, the value of $A$ has been
 determined to be $0.469\,(n_s/0.98)^{-0.35}\pm 0.017$. Here
 the scalar spectral index $n_s$ is taken to be $0.963$, which
 has been updated from the WMAP 7-year (WMAP7) data~\cite{r31}.
 Now, the total $\chi^2=\tilde{\chi}^2_\mu+\chi^2_A$,
 where $\tilde{\chi}^2_\mu$ is given in Eq.~(\ref{eq19}), and
 $\chi^2_A=(A-A_{obs})^2/\sigma_A^2$. Again, we scan the
 $\Omega_{m0}-s$ parameter space, and find that the best fit
 has $\chi^2_{min}=562.227$, and the corresponding
 best-fit parameters are $\Omega_{m0}=0.278$ and $s=-0.008$.
 In Fig.~\ref{fig2}, we present the corresponding $68.3\%$ and
 $95.4\%$ C.L. contours in the $\Omega_{m0}-s$ parameter space.
 Comparing with Fig.~\ref{fig1}, it is easy to see that the
 contours are significantly shrunk. Although the best-fit $s$
 is very close to zero, there is still a room for
 a significantly non-zero $s$ in the $95.4\%$~C.L. region. In
 fact, the viable $s$ can extend to about $-1.05$
 at $95.4\%$~C.L., or $-0.6$ at $68.3\%$~C.L.

%============================= Fig. 3 =================================

 \begin{center}
 \begin{figure}[tbhp]
 \centering
 \includegraphics[width=0.5\textwidth]{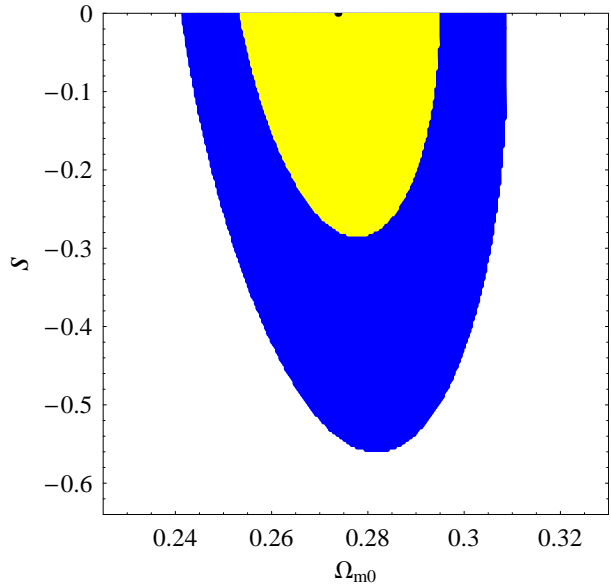}
 \caption{\label{fig3}
 The same as in Fig.~\ref{fig1}, except that this result is
 obtained by using the combined observational data of 580
 Union2.1 SNIa, the distance parameter $A$ from BAO, and the
 shift parameter $R$ from WMAP7.}
 \end{figure}
 \end{center}

%======================================================================

\vspace{-6mm} % used here just for a more comfortable typesetting

Finally, we further add the data from the observation of the
 cosmic microwave background (CMB). Here we use the the shift
 parameter $R$, which contains the main information of the
 observations of the CMB~\cite{r31,r32,r33}. The shift
 parameter $R$ of the CMB is defined by~\cite{r32,r33}
 \be{eq21}
 R\equiv\Omega_{m0}^{1/2}\int_0^{z_\ast}
 \frac{d\tilde{z}}{E(\tilde{z})}\,,
 \ee
 where the redshift of recombination $z_\ast=1091.3$
 which has been updated in the WMAP7 data~\cite{r31}. The shift
 parameter $R$ relates the angular diameter distance to
 the last scattering surface, the comoving size of the sound
 horizon at $z_\ast$ and the angular scale of the first
 acoustic peak in CMB power spectrum of temperature
 fluctuations~\cite{r32,r33}. The value of $R$ has been updated
 to $1.725\pm 0.018$ from the WMAP7 data~\cite{r31}. Now, the
 total $\chi^2=\tilde{\chi}^2_\mu+\chi^2_A+\chi^2_R$, where
 $\chi^2_R=(R-R_{obs})^2/\sigma_R^2$. We scan the
 $\Omega_{m0}-s$ parameter space, and find that the best fit
 has $\chi^2_{min}=562.546$, and the corresponding
 best-fit parameters are $\Omega_{m0}=0.274$ and $s=0.0$.
 In Fig.~\ref{fig3}, we present the corresponding $68.3\%$ and
 $95.4\%$ C.L. contours in the $\Omega_{m0}-s$ parameter space.
 Comparing with Fig.~\ref{fig2}, it is easy to see that the
 contours are further shrunk. Although the best-fit $s$
 becomes zero, there is still a room for a non-zero $s$ in the
 $95.4\%$~C.L. region. In fact, the viable $s$ can extend to
 about $-0.56$ at $95.4\%$~C.L., or $-0.29$ at $68.3\%$~C.L.

%============================= section 4 ===================================

\section{Concluding remarks}\label{sec4}

Some remarks are in order. Firstly, as mentioned above, pilgrim
 dark energy model reduces to the well-known $\Lambda$CDM model
 if $s=0$. On the other hand, when we constrain pilgrim dark
 energy by using the latest observational data, the best-fit
 parameter $s$ is zero. However, this does not mean that
 pilgrim dark energy fails, because there is still a room for
 a non-zero $s$ in the $95.4\%$~C.L. region. For instance,
 pilgrim dark energy with $s=-1/2$ is still viable. On the
 other hand, as is well known, $\Lambda$CDM model is plagued
 with the cosmological constant problem (the fine-tuning
 problem) and the cosmological coincidence problem. Therefore,
 pilgrim dark energy still deserves further investigations.

Secondly, although $w_\Lambda<-1$ in the whole cosmic history,
 it will go asymptotically to $-1$ in the late time. So, our
 universe will end in a de Sitter phase, rather than big rip.
 This is an advantage of pilgrim dark energy model.

Thirdly, in the present work, we have chosen IR cut-off
 $L=H^{-1}$ for pilgrim dark energy, since the Hubble horizon
 $H^{-1}$ is the simplest case. However, one can instead choose
 other IR cut-offs $L$ for pilgrim dark energy, such as the
 particle horizon $R_H\equiv a\int_0^t d\tilde{t}/a=a\int_0^a
 d\tilde{a}/(H\tilde{a}^2)$~\cite{r7}, the future
 event horizon $R_h\equiv a\int_t^\infty d\tilde{t}/a
 =a\int_a^\infty d\tilde{a}/(H\tilde{a}^2)$~\cite{r8}, the
 Ricci scalar curvature radius, which is actually proportional
 to the causal connection scale of perturbations in the flat
 universe $R_{\rm CC}=(\dot{H}+2H^2)^{-1/2}$~\cite{r9}, the
 formal generalization of $R_{\rm CC}$, namely
 $(\alpha H^2+\beta\dot{H})^{-1/2}$~\cite{r10}, the age of our
 universe $T=\int_0^a d\tilde{a}/(H\tilde{a})$~\cite{r11},
 the conformal age of our universe $\eta\equiv\int_0^t
 d\tilde{t}/a=\int_0^a d\tilde{a}/(\tilde{a}^2 H)$~\cite{r12},
 the radius of the cosmic null hypersurface~\cite{r13}, the
 so-called conformal-age-like length~\cite{r14}. This situation
 is very similar to holographic dark energy. It is possible
 to have completely new results (e.g. the cosmological
 constraints might be different) if $L$ is changed. Of course,
 we welcome other authors to explore this possibility.

Fourthly, as is shown above, pilgrim dark energy is based
 on the aforementioned speculation that the repulsive force
 contributed by the phantom-like dark energy ($w<-1$) is
 strong enough to prevent the formation of the black hole.
 In fact, this speculation is proposed just from
 some arguments, and it has no solid foundation so far. If
 one cannot agree the physical motivation of pilgrim dark
 energy presented here, we suggest that one can instead
 regard pilgrim dark energy as a purely phenomenological
 model without invoking any physical motivation.

Finally, if we consider pilgrim dark energy as a purely
 phenomenological model without any physical motivation, the
 physical requirement $s\leq 0$ can be given up. So, pilgrim
 dark energy can have a richer phenomenology. On the other
 hand, we also note that pilgrim dark energy can include
 other existing dark energy models as its special cases.
 For instance, if $s=1$ and $L=H^{-1}$, pilgrim
 dark energy reduces to the so-called QCD ghost
 dark energy~\cite{r34,r35,r36,r37},
 and Dvali-Gabadadze-Porrati (DGP) model~\cite{r38,r39}.
 Of course, if $s=2$ and choosing various $L$, pilgrim
 dark energy reduces to holographic dark energy~\cite{r8},
 (new)~agegraphic dark energy~\cite{r11,r12}, Ricci dark
 energy~\cite{r9}, and so on. If $L=H^{-1}$ and $s$ is
 free, pilgrim dark energy reduces to the modified DGP
 model ($\alpha$ dark energy)~\cite{r40}, and the modified
 holographic dark energy with IR infinite extra
 dimensions~\cite{r41}. If $L=H^{-1}$, $R_H$, $R_h$, and
 $s$ is free, pilgrim dark energy reduces to the so-called
 holographic cosmological ``constant'' derived
 from a generalized holographic relation between UV and
 IR cut-offs~\cite{r42}. We anticipate that pilgrim dark
 energy with various $L$ and a~completely free parameter $s$
 (which is not necessary to be $s\leq 0$) is a rich mine.

%============================= acknowledgements =================================

\section*{ACKNOWLEDGEMENTS}
We are grateful to Professors Rong-Gen~Cai and Shuang~Nan~Zhang
 for helpful discussions. We~also thank Minzi~Feng, as well as
 Yun-Song~Piao, Chao-Jun~Feng, Taotao~Qiu, Long-Fei~Wang and
 Xiao-Jiao~Guo, for kind help and discussions. This work was
 supported in part by NSFC under Grants No.~11175016 and
 No.~10905005, as well as NCET under Grant No.~NCET-11-0790,
 and the Fundamental Research Fund of Beijing Institute
 of Technology.

\renewcommand{\baselinestretch}{1.1}

%============================= references ==================================


\begin{thebibliography}{99}

\bibitem{r1}
E.~J.~Copeland, M.~Sami and S.~Tsujikawa,
 Int.\ J.\ Mod.\ Phys.\  D {\bf 15}, 1753 (2006) [hep-th/0603057];\\
J.~Frieman, M.~Turner and D.~Huterer,
 Ann.\ Rev.\ Astron.\ Astrophys.\  {\bf 46}, 385 (2008)
 [arXiv:0803.0982];\\
S.~Tsujikawa, arXiv:1004.1493 [astro-ph.CO];\\
D.~H.~Weinberg {\it et al.}, arXiv:1201.2434 [astro-ph.CO].

\bibitem{r2}
J.~D.~Bekenstein, Phys.\ Rev.\  D {\bf 7} (1973) 2333;\\
J.~D.~Bekenstein, Phys.\ Rev.\  D {\bf 9}, 3292 (1974);\\
J.~D.~Bekenstein, Phys.\ Rev.\  D {\bf 23}, 287 (1981);\\
J.~D.~Bekenstein,
 Phys.\ Rev.\  D {\bf 49}, 1912 (1994) [gr-qc/9307035].

\bibitem{r3}
S.~W.~Hawking,
 Commun.\ Math.\ Phys.\  {\bf 43}, 199 (1975)
 [Erratum-ibid.\  {\bf 46}, 206 (1976)];\\
S.~W.~Hawking, Phys.\ Rev.\  D {\bf 13}, 191 (1976).

\bibitem{r4}
A.~G.~Cohen, D.~B.~Kaplan and A.~E.~Nelson,
 Phys.\ Rev.\ Lett.\  {\bf 82}, 4971 (1999) [hep-th/9803132].

\bibitem{r5}
P.~Horava and D.~Minic,
 Phys.\ Rev.\ Lett.\  {\bf 85}, 1610 (2000) [hep-th/0001145];\\
S.~D.~Thomas, Phys.\ Rev.\ Lett.\  {\bf 89}, 081301 (2002).

\bibitem{r6}
S.~D.~H.~Hsu,
 Phys.\ Lett.\  B {\bf 594}, 13 (2004) [hep-th/0403052].

\bibitem{r7}
W.~Fischler and L.~Susskind, hep-th/9806039;\\
R.~Bousso, JHEP {\bf 9907}, 004 (1999) [hep-th/9905177].

\bibitem{r8}
M.~Li, Phys.\ Lett.\  B {\bf 603}, 1 (2004) [hep-th/0403127].

\bibitem{r9}
C.~J.~Gao, X.~L.~Chen and Y.~G.~Shen,
 Phys.\ Rev.\  D {\bf 79}, 043511 (2009) [arXiv:0712.1394];\\
R.~G.~Cai, B.~Hu and Y.~Zhang,
 Commun.\ Theor.\ Phys.\  {\bf 51}, 954 (2009)
 [arXiv:0812.4504].

\bibitem{r10}
L.~N.~Granda and A.~Oliveros,
 Phys.\ Lett.\  B {\bf 669}, 275 (2008) [arXiv:0810.3149];\\
L.~N.~Granda and A.~Oliveros,
 Phys.\ Lett.\  B {\bf 671}, 199 (2009) [arXiv:0810.3663].

\bibitem{r11}
R.~G.~Cai,
 Phys.\ Lett.\  B {\bf 657}, 228 (2007) [arXiv:0707.4049];\\
H.~Wei and R.~G.~Cai,
 Eur.\ Phys.\ J.\  C {\bf 59}, 99 (2009) [arXiv:0707.4052];\\
H.~Wei and R.~G.~Cai,
 Phys.\ Lett.\  B {\bf 655}, 1 (2007) [arXiv:0707.4526].

\bibitem{r12}
H.~Wei and R.~G.~Cai,
 Phys.\ Lett.\  B {\bf 660}, 113 (2008) [arXiv:0708.0884];\\
H.~Wei and R.~G.~Cai,
 Phys.\ Lett.\  B {\bf 663}, 1 (2008) [arXiv:0708.1894].

\bibitem{r13}
C.~J.~Gao, arXiv:1108.5827 [gr-qc].

\bibitem{r14}
Z.~P.~Huang and Y.~L.~Wu, arXiv:1202.2590 [hep-th];\\
Z.~P.~Huang and Y.~L.~Wu, arXiv:1202.3517 [astro-ph.CO];\\
Y.~Ling and W.~J.~Pan, arXiv:1205.0209 [gr-qc].

\bibitem{r15}
E.~Babichev, V.~Dokuchaev and Y.~Eroshenko,
 Phys.\ Rev.\ Lett.\  {\bf 93}, 021102 (2004) [gr-qc/0402089];\\
E.~Babichev, V.~Dokuchaev and Y.~Eroshenko,
 J.\ Exp.\ Theor.\ Phys.\  {\bf 100}, 528 (2005)
 [astro-ph/0505618];\\
E.~Babichev, V.~Dokuchaev and Y.~Eroshenko, gr-qc/0507119.

\bibitem{r16}
C.~J.~Gao, X.~L.~Chen, V.~Faraoni and Y.~G.~Shen,
 Phys.\ Rev.\  D {\bf 78}, 024008 (2008) [arXiv:0802.1298].

\bibitem{r17}
J.~A.~Gonzalez and F.~S.~Guzman,
 Phys.\ Rev.\  D {\bf 79}, 121501 (2009) [arXiv:0903.0881].

\bibitem{r18}
C.~Y.~Sun,
 Commun.\ Theor.\ Phys.\  {\bf 52}, 441 (2009) [arXiv:0812.2996].

\bibitem{r19}
M.~Jamil and A.~Qadir,
 Gen.\ Rel.\ Grav.\  {\bf 43}, 1069 (2011) [arXiv:0908.0444].

\bibitem{r20}
M.~Sharif and G.~Abbas,
 Chin.\ Phys.\ Lett.\  {\bf 28}, 090402 (2011) [arXiv:1109.1043];\\
M.~Sharif and G.~Abbas,
 Chin.\ Phys.\ Lett.\  {\bf 29}, 010401 (2012) [arXiv:1202.1705].

\bibitem{r21}
T.~Harada, H.~Maeda and B.~J.~Carr,
 Phys.\ Rev.\  D {\bf 74}, 024024 (2006) [astro-ph/0604225].

\bibitem{r22}
G.~Chapline, ``Dark energy stars'', in the Proceedings of 22nd
 Texas Symposium on Relativistic Astrophysics at Stanford
 University, Stanford, California, 13-17 Dec 2004, pp 0205
 [astro-ph/0503200].

\bibitem{r23}
Z.~H.~Li and A.~Z.~Wang,
 Mod.\ Phys.\ Lett.\  A {\bf 22}, 1663 (2007) [astro-ph/0607554].

\bibitem{r24}
F.~Rahaman, A.~Ghosh and M.~Kalam,
 Nuovo Cim.\  B {\bf 121}, 279 (2006) [gr-qc/0612154].

\bibitem{r25}
N.~Suzuki {\it et al.} [SCP Collaboration],
 Astrophys.\ J.\  {\bf 746}, 85 (2012) [arXiv:1105.3470];\\
 The numerical data of the full Union2.1 sample
 are available at
 http:$/\!/$supernova.lbl.gov/Union

\bibitem{r26}
E.~Di Pietro and J.~F.~Claeskens,
 Mon.\ Not.\ Roy.\ Astron.\ Soc.\  {\bf 341}, 1299 (2003)
 [astro-ph/0207332].

\bibitem{r27}
S.~Nesseris and L.~Perivolaropoulos,
 Phys.\ Rev.\ D {\bf 72}, 123519 (2005) [astro-ph/0511040];\\
L.~Perivolaropoulos,
 Phys.\ Rev.\ D {\bf 71}, 063503 (2005) [astro-ph/0412308].

\bibitem{r28}
H.~Wei, X.~J.~Guo and L.~F.~Wang,
 Phys.\ Lett.\ B {\bf 707}, 298 (2012) [arXiv:1112.2270];\\
H.~Wei, JCAP {\bf 1104}, 022 (2011) [arXiv:1012.0883];\\
H.~Wei,
 Commun.\ Theor.\ Phys.\  {\bf 56}, 972 (2011) [arXiv:1010.1074];\\
H.~Wei,
 Phys.\ Lett.\  B {\bf 692}, 167 (2010) [arXiv:1005.1445];\\
H.~Wei, JCAP {\bf 1008}, 020 (2010) [arXiv:1004.4951];\\
H.~Wei,
 Phys.\ Lett.\  B {\bf 691}, 173 (2010) [arXiv:1004.0492];\\
H.~Wei,
 Phys.\ Lett.\  B {\bf 687}, 286 (2010) [arXiv:0906.0828];\\
H.~Wei,
 Eur.\ Phys.\ J.\  C {\bf 62}, 579 (2009) [arXiv:0812.4489];\\
H.~Wei,
 Eur.\ Phys.\ J.\  C {\bf 60}, 449 (2009) [arXiv:0809.0057];\\
H.~Wei and S.~N.~Zhang,
 Eur.\ Phys.\ J.\  C {\bf 63}, 139 (2009) [arXiv:0808.2240].

\bibitem{r29}
M.~Tegmark {\it et al.} [SDSS Collaboration],
 Phys.\ Rev.\ D {\bf 69}, 103501 (2004) [astro-ph/0310723];\\
M.~Tegmark {\it et al.} [SDSS Collaboration],
 Astrophys.\ J.\  {\bf 606}, 702 (2004) [astro-ph/0310725];\\
U.~Seljak {\it et al.} [SDSS Collaboration],
 Phys.\ Rev.\ D {\bf 71}, 103515 (2005) [astro-ph/0407372];\\
M.~Tegmark {\it et al.} [SDSS Collaboration],
 Phys.\ Rev.\  D {\bf 74}, 123507 (2006) [astro-ph/0608632].

\bibitem{r30}
D.~J.~Eisenstein {\it et al.} [SDSS Collaboration],
 Astrophys.\ J.\  {\bf 633}, 560 (2005) [astro-ph/0501171].

\bibitem{r31}
E.~Komatsu {\it et al.} [WMAP Collaboration],
 Astrophys.\ J.\ Suppl.\  {\bf 192}, 18 (2011) [arXiv:1001.4538].

\bibitem{r32}
Y.~Wang and P.~Mukherjee,
 Astrophys.\ J.\  {\bf 650}, 1 (2006) [astro-ph/0604051].

\bibitem{r33}
J.~R.~Bond, G.~Efstathiou and M.~Tegmark,
 Mon.\ Not.\ Roy.\ Astron.\ Soc.\  {\bf 291}, L33 (1997)
 [astro-ph/9702100].

\bibitem{r34}
F.~R.~Urban and A.~R.~Zhitnitsky,
 Phys.\ Lett.\  B {\bf 688}, 9 (2010) [arXiv:0906.2162];\\
F.~R.~Urban and A.~R.~Zhitnitsky,
 Phys.\ Rev.\  D {\bf 80}, 063001 (2009) [arXiv:0906.2165];\\
F.~R.~Urban and A.~R.~Zhitnitsky,
 JCAP {\bf 0909}, 018 (2009) [arXiv:0906.3546];\\
F.~R.~Urban and A.~R.~Zhitnitsky,
 Nucl.\ Phys.\  B {\bf 835}, 135 (2010) [arXiv:0909.2684].

\bibitem{r35}
N.~Ohta,
 Phys.\ Lett.\  B {\bf 695}, 41 (2011) [arXiv:1010.1339].

\bibitem{r36}
R.~G.~Cai, Z.~L.~Tuo, H.~B.~Zhang and Q.~Su,
 Phys.\ Rev.\  D {\bf 84}, 123501 (2011) [arXiv:1011.3212];\\
R.~G.~Cai, Z.~L.~Tuo, Y.~B.~Wu and Y.~Y.~Zhao,
 arXiv:1201.2494 [astro-ph.CO].

\bibitem{r37}
C.~J.~Feng, X.~Z.~Li and X.~Y.~Shen, arXiv:1202.0058 [astro-ph.CO];\\
C.~J.~Feng, X.~Z.~Li and X.~Y.~Shen, arXiv:1105.3253 [hep-th].

\bibitem{r38}
C.~Deffayet,
 Phys.\ Lett.\  B {\bf 502}, 199 (2001) [hep-th/0010186];\\
C.~Deffayet, G.~R.~Dvali and G.~Gabadadze,
 Phys.\ Rev.\  D {\bf 65}, 044023 (2002) [astro-ph/0105068].

\bibitem{r39}
A.~Lue, R.~Scoccimarro and G.~D.~Starkman,
 Phys.\ Rev.\  D {\bf 69}, 124015 (2004) [astro-ph/0401515];\\
A.~Lue,
 Phys.\ Rept.\  {\bf 423}, 1 (2006) [astro-ph/0510068].

\bibitem{r40}
G.~Dvali and M.~S.~Turner, astro-ph/0301510.

\bibitem{r41}
Y.~G.~Gong and T.~J.~Li,
 Phys.\ Lett.\  B {\bf 683}, 241 (2010) [arXiv:0907.0860].

\bibitem{r42}
C.~J.~Feng,
 Phys.\ Lett.\  B {\bf 663}, 367 (2008) [arXiv:0709.2456].

\bibitem{r43}
S.~Nojiri and S.~D.~Odintsov,
 Gen.\ Rel.\ Grav.\  {\bf 38}, 1285 (2006) [hep-th/0506212].

\bibitem{r44}
R.~Akhoury, C.~S.~Gauthier and A.~Vikman,
 JHEP {\bf 0903}, 082 (2009) [arXiv:0811.1620].

\end{thebibliography}
\end{document}